\documentclass[conference]{IEEEtran}
\IEEEoverridecommandlockouts
\usepackage{cite}
\usepackage{amsmath,amssymb,amsfonts}
\usepackage{algorithmic}
\usepackage{graphicx}
\usepackage{textcomp}
\usepackage{xcolor}
\def\BibTeX{{\rm B\kern-.05em{\sc i\kern-.025em b}\kern-.08em
    T\kern-.1667em\lower.7ex\hbox{E}\kern-.125emX}}

\usepackage{svg}
\usepackage{graphicx} 
\usepackage{caption}
\usepackage{subcaption}
\usepackage{hyperref}
\usepackage{booktabs}

\newcommand{\blue}[1]{\textcolor{blue}{#1}}

\begin{document}

\title{Serverless Query Processing with Flexible Performance SLAs and Prices
}

\author{
\IEEEauthorblockN{Haoqiong Bian, Dongyang Geng, Panfeng Guo, Yunpeng Chai}
\IEEEauthorblockA{Renmin University of China\\
\{bianhq, gengdongyang, panfengguo, ypchai\}@ruc.edu.cn}
\and
\IEEEauthorblockN{Anastasia Ailamaki}
\IEEEauthorblockA{EPFL\\
anastasia.ailamaki@epfl.ch}
}

\maketitle

\begin{abstract}
Serverless query processing has become increasingly popular due to its auto-scaling, high elasticity, and pay-as-you-go pricing.
It allows cloud data warehouse (or lakehouse) users to focus on data analysis without the burden of managing systems and resources.
Accordingly, in serverless query services, users become more concerned about cost-efficiency under acceptable performance than performance under fixed resources.
This poses new challenges for serverless query engine design in providing flexible performance service-level agreements (SLAs) and cost-efficiency (i.e., prices).

In this paper, we first define the problem of flexible performance SLAs and prices in serverless query processing and discuss its significance.
Then, we envision the challenges and solutions for solving this problem and the opportunities it raises for other database research.
Finally, we present PixelsDB, an open-source prototype with three service levels supported by dedicated architectural designs.
Evaluations show that PixelsDB reduces resource costs by 65.5\% for near-real-world workloads generated by Cloud Analytics Benchmark (CAB) while not violating the pending time guarantees.
\end{abstract}


\section{Introduction}\label{sec:intro}

Serverless query processing, also known as Query-as-a-Service (QaaS), has become the new paradigm of analytical query processing in the cloud.
In contrast to traditional cloud data warehouses, serverless query services such as Athena~\cite{aws-athena} and BigQuery~\cite{google-bigquery} allow users to run queries without considering the system setup and resource quota.
They can quickly allocate computing units for queries and scale to zero when idle.
Meanwhile, users are billed on a per-query basis according to the consumed computing units~\cite{aws-redshift-serverless-billing} or scanned data size~\cite{aws-athena-price,google-bigquery}.
These benefits significantly improve the usability of the data warehouses, freeing users from the burden of system operations and resource management.
Accordingly, unlike in traditional cloud data warehouses where users focus on query performance under fixed resources, users of serverless query services become more concerned about cost-efficiency under acceptable performance~\cite{cost-intelligent-data-analysis,resource-adaptive-query-execution}.
Improving such user-observable cost-efficiency is orthogonal to reducing the overall costs of multi-tenancy in a cloud data warehouse~\cite{cost-intelligent-data-analysis}.

An ideal 
solution for this problem is to design a query engine that provides Pareto-optimal cost-efficiency and performance for each query.
However, this is very challenging as it requires the query engine to calculate and apply the optimal settings adaptively for any query under any execution time and budget constraints~\cite{cost-intelligent-data-analysis}.
Fortunately, in real-world applications, queries are often naturally classified according to scenarios~\cite{cloud-analytics-benchmark,redshift-redset}.
For example, we may have frequent repeating queries from busy dashboards, interactive queries for ad-hoc analytics, and daily queries for reports in the same organization.
Users typically have different preferences between performance and price for different queries.
For example, users may prefer higher performance for interactive queries but lower costs for daily report queries.

With this observation, we can simplify the problem to providing enumerable levels of performance and prices for users to flexibly choose from on a per-query basis (flexible SLA problem, for short).
To verify that flexible SLA really addresses users' concerns about cost-efficiency, we conducted a user study among database practitioners on the pricing preferences of cloud data warehouses.
In the 109 valid questionnaires collected, 72.5\% of users prefer having flexible SLA in cloud data warehouses.
This ratio is especially high among database researchers/developers and experienced database users, reaching 88.2\% and 81.3\%, respectively.
Note that flexible SLA is not simply a pricing problem.
The key point is to ensure lower resource consumption for a query under more relaxed performance guarantees.
This is beneficial to both users and cloud vendors.

In this paper, we begin by summarizing the architectures of existing serverless query services.
Then, after defining the flexible SLA problem and verifying its significance through the user study, we envision the architectural designs to implement flexible SLA and the opportunities it brings to database research.
We propose that a stage-oriented scaling (SOS) architecture can efficiently support flexible SLAs.
In this architecture, each stage in the query plan DAG runs independently in a serverless computing environment optimized for SOS, and each computing unit has isolated resources to run a simple task on a data split.
This makes query execution more deterministic and controllable.
The main challenge is to reduce the performance and resource overhead caused by the isolated computing units.
Following this architectural design, we present PixelsDB, an open-source (\blue{\href{https://github.com/pixelsdb/pixels}{https://github.com/pixelsdb/pixels}}) query and storage engine for serverless data analytics.
PixelsDB currently supports three service levels regarding query pending time:
(1) \textit{Immediate} that starts executing users' queries immediately;
(2) \textit{Relaxed} that guarantees users' queries start executing within a certain time (e.g., 5 mins);
and (3) \textit{Best-of-effort (BoE)} that executes queries at best effort without clear guarantees.
Evaluations on near-real-world workloads generated by Cloud Analytics Benchmark (CAB)~\cite{cloud-analytics-benchmark} show that while guaranteeing the query pending time, the costs of relaxed and BoE queries can be reduced by up to 64.5\% and 95\%, respectively, without significantly affecting query execution time.

\section{Serverless Query Processing}\label{sec:serverless-query-processing}

In this section, we discuss the architectural design of state-of-the-art serverless query services, which provides the preliminaries for the discussions and system design in Sections~\ref{sec:flexible-sla} and~\ref{sec:pixelsdb-design}.

\subsection{Micro-servers}\label{subsec:micro-servers}
Serverless indicates that the servers are invisible to users.
However, servers still exist under the hood.
Existing serverless query services are generally backed by virtual micro-servers, which are enabled by lightweight virtualization technologies such as containerization and micro-virtual-machine (micro-VM).
For example, Athena for Spark uses Firecracker micro-VMs~\cite{firecracker-paper,athena-for-spark-uses-firecracker}, whereas BigQuery uses the fine-grained containerized units provided by Borg~\cite{google-borg-paper,dremel-2020}.

\textbf{Why micro-servers?}
Serverless query services have two fundamental requirements that differ from traditional query engines: 
(1) fine-grained resource allocation and billing on a per-query basis;
and (2) automatic and elastic resource scaling for workload changes.
Virtual micro-servers provide the resource management flexibility to meet these requirements:
(1) Micro-servers can be efficiently created and released according to load changes~\cite{overview-of-bigquery-architecture};
(2) Micro-servers well-suit the heterogeneous physical servers that are common in large cloud computing platforms~\cite{google-borg-paper};
(3) Allocating micro-servers for each query simplifies per-query price calculation~\cite{aws-redshift-serverless-billing,aws-athena-price,google-bigquery-price}.

Commercial serverless query services may expose abstract resource units other than micro-servers to describe the concepts in resource management for customers.
For example, Redshift-serverless uses \textit{Redshift Processing Unit (RPU)} to describe the resource capacity~\cite{aws-redshift-serverless-billing}.
Each RPU provides 16 GB memory~\cite{redshift-serverless-considerations}.
Athena has a similar concept called \textit{Data Processing Unit (DPU)}, which provides 4 vCPUs and 16 GB memory~\cite{aws-athena-price}.

\subsection{Architectural Principles}\label{subsec:serverless-arch-principles}

Despite micro-servers, serverless query processing applies the following two common design principles.

\textbf{Disaggregated compute and storage.}
Similar to other cloud databases, serverless query services generally decouple compute and storage into two layers~\cite{aws-athena-paper,redshift-reinvented,overview-of-bigquery-architecture}.
The storage layer of serverless query services is composed of cloud object storage (e.g., S3~\cite{aws-s3}), the open file formats (e.g., Parquet~\cite{parquetMainPage} and ORC~\cite{ORCMainPage}), and the catalog manager (e.g., Glue~\cite{aws-glue}).
Whereas the compute layer is mainly composed of the query engine, the scaling manager, and the micro-server computing environment.
Such disaggregated architecture allows computing and storage resources to scale independently, improving the elasticity and utilization of resources.

\textbf{In-situ data analytics.}
With disaggregated compute and storage, the compute nodes no longer manage user data.
Serverless query services allow analytical applications to easily register the open-format files as tables and perform in-situ data analysis without expensive ETL.
This improves the timeliness of data analytics.

\subsection{Scaling Paradigms}\label{subsec:resource-scaling-paradigms}
Based on the above principles, we further discuss how serverless query services scale computing resources to process queries.

\textbf{Plan-oriented Scaling (POS).}
In some serverless query services, the query engine in micro-servers inherits the implementation of existing MPP systems.
For example, the query engine in Amazon Athena was forked from Presto~\cite{presto-paper} (a lightweight MPP system), and it periodically merges updates from the main branches of Presto and Trino~\cite{trino-main-page}~\cite{aws-athena-v1-v2,aws-athena-v3}.
Redshift-serverless reuses the MPP query engine of Redshift and enhances its auto-tuning capabilities~\cite{redshift-reinvented}.
Although we can hardly study the detailed design of these query engines, they should have inherited the MPP-style scaling paradigm:
horizontally divide the query plan into parallel tasks and execute these tasks in a horizontally scalable compute cluster.
Each compute node in the cluster is fully functional for executing the query plan on a data split.
We call this scaling paradigm plan-oriented scaling (POS), as the query plan is the object on which scaling acts.

\textbf{Stage-oriented Scaling (SOS).}
A query plan is a DAG of physical stages.
Stage-oriented scaling (SOS) differs from POS as:
each stage in the query plan is executed in a set of parallel compute instances (namely \textit{workers}).
Each worker is responsible for executing a single stage on a data split.
For example, a \textit{hash-partition worker} may read a data split from upstream and do hash partitioning.
SOS is used in commercial systems such as BigQuery (Dremel)~\cite{dremel-2020,overview-of-bigquery-architecture} and academic systems such as Starling~\cite{starling-paper} and Pixels-Turbo~\cite{pixels-turbo}.

SOS requires the computing environment to provide a simple and reliable interface for developing, deploying, and running the query stage tasks.
For example, BigQuery (Dremel) implements SOS on Google's containerized job computing environment Borg~\cite{google-borg-paper}.
Whereas Starling, Lambada, and Pixels-Turbo implement SOS on cloud function (CF) services such as AWS Lambda~\cite{aws-lambda}.


\section{Flexible SLA}\label{sec:flexible-sla}

In this section, we first define the flexible SLA problem and verify its significance through a user study among database practitioners.
Then, we envision the architectural design to address this problem and the opportunities it brings to database research.

\subsection{Problem Definition}\label{subsec:problem-definition}

As discussed in Section~\ref{sec:serverless-query-processing}, existing serverless query services have reduced the burden of resource and system management for users.
They allow users to submit queries to a service endpoint and pay on-demand.
On this basis, users are paying more attention to cost-efficiency under acceptable performance~\cite{cost-intelligent-data-analysis,resource-adaptive-query-execution}.
This is in line with human nature.
When people can buy something on-demand, they would prefer to make a choice based on the quality (or performance) and price, just like how people buy apples from a fruit shop.

However, users' preferences for performance and price are dynamic.
Therefore in~\cite{cost-intelligent-data-analysis}, the problem is defined as providing Pareto-optimal cost-efficiency and performance for queries; thus, given any performance SLA and/or monetary budget of a query, the query engine can calculate and apply the optimal settings.
This problem is very challenging for system design and implementation, although~\cite{cost-intelligent-data-analysis} proposes reasonable sketches to address it.

In this paper, we find this problem can be simplified in real scenarios.
We observe that in real-world applications, queries are often naturally classified~\cite{cloud-analytics-benchmark,redshift-redset}, and every query type may have different preferences for performance and price.
For example, we may have repeating dashboard queries, interactive ad-hoc queries, and daily or weekly report queries in the same organization.
Among these query types, users may prefer higher performance for interactive and dashboard queries but lower costs for report queries.
Intuitively, we can meet users' preferences by providing individual performance-price trade-offs for each query type.

\textbf{The flexible SLA problem.}
With this observation, this paper defines a simplified form for the problem:
Providing enumerable levels of performance and price guarantees for users to choose from for each query type flexibly.
The performance and price guarantees can be either \textbf{absolute} (e.g., the query latency is $\leq$ 10 seconds, and the query cost is $\leq$ 50 cents) or \textbf{relative} (e.g., compared to another service level, the performance and cost of each query is 30\% and 50\% lower, respectively).
We call this problem the \textit{flexible SLA} problem. 

We believe relative guarantees (i.e., SLA) are good enough for users in most conditions.
With relative service levels, the query engine does not need to control the latency and costs of each query accurately.
It only needs to ensure that a query generally consumes fewer (or cheaper) resources at a lower performance SLA.

\subsection{A User Study}\label{subsec:user-study}
To verify that flexible SLA can indeed address users' growing demands for controllable cost and performance, we conducted a user study among database users, developers, and researchers.
This user study uses a questionnaire with eight questions.

\textbf{Q1 and Q2 are to determine the profile of the participants.}
We ask if they best fit (1) database researchers and developers, (2) database users, (3) students who learned databases, or (4) others.
We also ask them how familiar they are with cloud data warehouses.

\textbf{Q3 is to filter out invalid questionnaires.}
We ask the participants to select cloud data warehouses from well-known database products (MySQL, Redis, Snowflake, MongoDB, and Redshift).

\textbf{Q4 and Q5 are to see users' preferences for pricing models.}
We ask the participants to give their preferences among provisioned pricing~\footnote{Provisioned pricing is still available in today's serverless query services~\cite{aws-athena-price,aws-redshift-serverless-billing,google-bigquery-price}.}, on-demand pricing, and flexible SLA with relative performance and price guarantees.

\textbf{Q6 is to see users' demand for absolute performance-price.}
We ask the participants if the cloud data warehouse can estimate the absolute price of each query under different performance guarantees, whether they rate this feature as (1) useless, (2) somewhat useful (might try it), or (3) very useful (will definitely use it).

\textbf{Q7 is to see users' demand for cost visibility.}
We ask the participants if the cloud data warehouse allows users to view and analyze the monetary costs of historical queries, how they rate this feature given the same options as Q6.

\textbf{Q8 is to see users' demand for text-to-SQL.}
This is an additional question beyond the scope of this paper.


We sent the questionnaire to database practitioners through Tencent Survey~\cite{tencent-survey}.
It was viewed by 887 people and answered by 416, out of which 109 submissions are valid.
The questionnaire, validation rules, and valid and invalid submissions are available at 
\blue{\href{https://github.com/pixelsdb/cdw-user-study}{https://github.com/pixelsdb/cdw-user-study}}.

\begin{figure}
     \centering
     \begin{subfigure}[b]{0.155\textwidth}
         \centering
         \includegraphics[width=\textwidth]{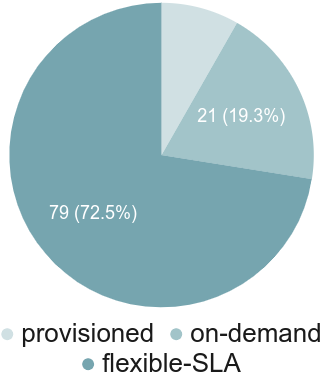}
         \vspace{-1em}
         \caption{Q4, Q5.}
         \label{subfig:q4-5}
     \end{subfigure}
     \begin{subfigure}[b]{0.15\textwidth}
         \centering
         \includegraphics[width=\textwidth]{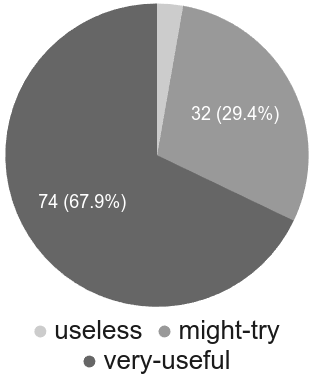}
         \vspace{-1em}
         \caption{Q6.}
         \label{subfig:q6}
     \end{subfigure}
     \begin{subfigure}[b]{0.15\textwidth}
         \centering
         \includegraphics[width=\textwidth]{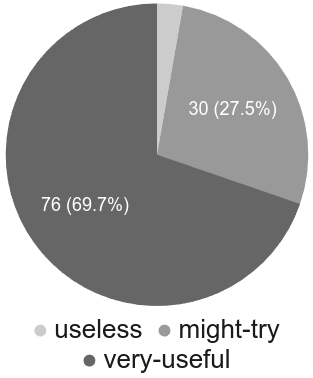}
         \vspace{-1em}
         \caption{Q7.}
         \label{subfig:q7}
     \end{subfigure}
     \vspace{-0.5em}
        \caption{Participants' preferences on pricing models (Q4, Q5), absolute performance-price (Q6), and cost visibility (Q7).}
        \label{fig:preferences-q4567}
        \vspace{-0em}
\end{figure}

\begin{figure}[tbp]
	\vspace{-0.5em}
	\begin{center}
		{\includegraphics[scale=0.36]{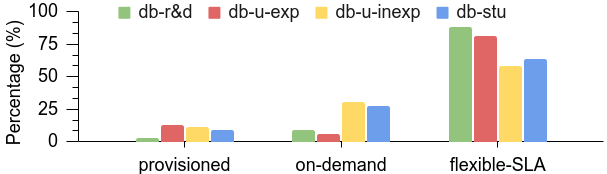}}
	\end{center}
	\vspace{-1em}
	\caption{\label{fig:pricing-model-preference}Pricing-model preferences per participant category.}
	\vspace{-0em}	
\end{figure}

In the valid submissions, there are 34 database researchers and developers (\textit{db-r\&d}), 59 database users, 11 students who learned databases (\textit{db-stu}), and 5 other participants.
Of the 59 database users, 16 are familiar with cloud data warehouses.
We call these users experienced database users (\textit{db-u-exp}) and the others inexperienced database users (\textit{db-u-inexp}).
Two main statistical patterns, SP1 and SP2, are shown in Figure~\ref{fig:preferences-q4567} and Figure~\ref{fig:pricing-model-preference}, respectively.


\textbf{SP1: Participants are comparably interested in relative flexible SLAs, absolute performance-price, and cost visibility.}
As shown in Figure~\ref{fig:preferences-q4567}, 72.5\%, 67.9\%, and 69.7\% of participants clearly chose to use relative flexible SLAs, absolute performance-price, and cost visibility, respectively.
This indicates that users are highly concerned about cost-efficiency, and flexible SLA can better address this concern than existing provisioned or on-demand pricing models.

\textbf{SP2: Senior practitioners prefer flexible SLAs more than junior ones.}
There are 50 senior practitioners (\textit{db-r\&d} and \textit{db-u-exp}) and 54 junior practitioners (\textit{db-u-inexp} and \textit{db-stu}).
As shown in Figure~\ref{fig:pricing-model-preference}, there are more senior practitioners who prefer flexible SLA than junior ones.
This further emphasizes the significance of flexible SLA in real-world analytical applications.

\subsection{Vision}\label{subsec:vision}
In fact, serverless query processing and flexible SLA can complement each other.
While flexible SLA improves the usability and competitiveness of serverless query services, the serverless paradigm transfers the control of resource quotas and system settings from users to cloud vendors, giving vendors the opportunity to implement flexible SLA.
For flexible SLA, we propose the following key visions.



\textbf{1. SOS is more promising for Flexible SLA.}
POS inherits the design and implementation of existing MPP systems, which are considered more performant and reliable~\cite{pixels-turbo,off-the-shelf-analytics-on-serverless}.
However, MPP (POS) is like a chaotic system.
Its implementation is very complex (e.g., most MPP engines have millions of lines of code in the kernel), and concurrent queries running in the same cluster may interfere with each other.
Although resource isolation solutions inside the MPP query engine (e.g., using cgroup) can reduce the resource contention between threads, the concurrent queries still compete for resources as the total amount of resources in a cluster is fixed in a certain period.
Creating a new MPP cluster or scaling out an existing MPP cluster often requires at least a few minutes.
For example, in our experimental evaluations, adding a new worker node into a Trino~\cite{trino-main-page} cluster running in EC2 typically requires more than 90 seconds for instance+os initialization plus more than 30 seconds for query engine startup.
Similar startup latencies are also shown by the evaluations in~\cite{off-the-shelf-analytics-on-serverless}.
Therefore, concurrent queries in a Trino cluster have to compete for the execution slots (aligned to the number of CPU cores) and available memory on each worker node.
Furthermore, the query operators in POS are also not isolated.
The operator pipelines in the same query plan may also compute for resources in the same worker node.
Hence, it is difficult 
to accurately predict and control a query's performance and cost in POS.
The cost model in the query optimizer can only estimate some basic performance metrics, such as the number of I/O blocks in a query.
Even reproducing an evaluation result would require running queries back-by-back in a specified hardware and software environment.

On the contrary, SOS runs simple stage tasks in isolated and lightweight micro-servers, reducing interference and software complexity.
As discussed in Section~\ref{subsec:resource-scaling-paradigms}, each stage task only executes a few basic operations.
In SOS query engines, such stage tasks can be implemented as functions that are invoked by the query coordinator~\cite{starling-paper,lambada-paper,pixels-turbo}.
In PixelsDB (see Section~\ref{sec:pixelsdb-design}), each function (stage task) only involves around 3-5K lines of code.
It is much easier to profile, predict, and control the performance and cost of the functions.
This makes SOS promising for flexible SLA.

We can also run POS in isolated mini-servers~\cite{off-the-shelf-analytics-on-serverless} and run each query in a separate POS cluster.
Thus, we avoid interference between concurrent queries.
However, since each node in the POS cluster has a complex code stack and maintains a lot of states, it is hard to scale the size of the cluster just in time when the query arrives.
We need to provision a POS cluster of the desired size according to the service level to support performance SLAs, like what Redshift-Serverless does~\cite{redshift-intelligent-scaling}.
This goes against the per-query resource allocation principle of serverless query processing.
In addition, each POS node often has a large fixed memory overhead to maintain many states and auxiliary data structures, resulting in low memory utilization in the mini-server.
For example, each Trino worker node requires at least 4-8GB of memory overhead.
Hence, in a mini-server with 16GB of memory, only 50-75\% of memory space is available for query execution.

With SOS, resource isolation and provision are provided by the underlying computing services, such as AWS ECS~\cite{aws-ecs} and Fargate~\cite{aws-fargate}.
This greatly simplifies the design of the query engine.
The query planner can estimate the amount of resources needed in each stage.
Given the simple code stack in each stage, it is easier to profile the performance characteristics of each stage and train machine learning models to predict execution time under different resource quotas and input sizes and distributions.


\textbf{2. SOS-optimized serverless computing will be the mainstream.}
The main problem of SOS is the (possibly) poor computing efficiency.
Most academic SOS query engines are based on cloud function (CF) services, which were not designed for data-intensive applications like query processing~\cite{aws-lambda,pixels-turbo,off-the-shelf-analytics-on-serverless}.
Whereas commercial systems such as Dremel~\cite{dremel-2020} and Borg~\cite{google-borg-paper} are not open for research.
CF has some limitations for query processing, such as:
(1) It does not ensure balanced network bandwidth among parallel compute instances~\cite{pixels-turbo};
(2) It cannot establish a dynamic query pipeline among compute instances~\cite{lambada-paper,pixels-turbo};
(3) Its CPU/memory ratio is fixed, which may cause internal resource fragmentation;
and (4) It does not have disaggregated memory like Dremel~\cite{dremel-2020}, making it inefficient to exchange intermediate results between stages.


However, given the unique advantages of SOS, we believe there will be serverless computing services optimized and become the mainstream for SOS query processing.
Just like how cloud object storage like S3~\cite{aws-s3} was designed for web apps but is now the de-facto standard for analytical storage.
For SOS-optimized serverless computing, it is vital to build a white-box computing framework for SOS based on existing cloud infrastructures (e.g., AWS ECS~\cite{aws-ecs} and Fargate~\cite{aws-fargate}). 
Thus, we can easily profile and evaluate the computing framework, as well as the SOS query engine on top of it.
Compared to on-premises systems, such a framework based on public IaaS reduces the complexity of implementation and operation, and it solves the problems of resource pricing and elasticity.
Research on this framework may inspire cloud vendors and feedback to public clouds.
We are implementing such a framework as discussed in Section~\ref{subsec:overview}.

\textbf{3. Flexible SLA provides a platform for database research on performance-price trade-offs.}
With SOS-based flexible SLA, the query engine will be more extendable due to the deep decoupling and isolation.
Such a query engine can become a platform for inspiring and integrating innovative ideas that trade performance for cost or vice versa.
Database technologies can be applied to implement service levels without significantly increasing system complexity.
Thus, the technologies are easier to be applied and find the target users.
For example, multi-query execution~\cite{multi-query-execution} can be applied to reduce query costs in the relaxed and BoE service levels discussed in Section~\ref{subsec:service-levels}.

\textbf{4. Automatic SLA Selection is important.}
With flexible SLAs, how to automatically identify users' performance-price requirements is an open problem.
Although users want to control performance and price (discussed in the user study in Section~\ref{subsec:user-study}) and have clear performance-price requirements for each type of query in many cases, requiring users to select a particular SLA for every query is still a cumbersome task.
Automatically selecting a default SLA when the user submits a query or giving a warning when the user mistakenly selects an inappropriate SLA will greatly improve the usability of a serverless query service.

\textbf{5. Absolute performance and price guarantees are necessary but challenging.}
As discussed in Section~\ref{subsec:problem-definition}, the performance and price guarantees of each service level can be relative or absolute.
We believe that relative guarantees are good enough in most cases.
This is shown by the user study in Section~\ref{subsec:user-study}.
79\% participants prefer the relative flexible-SLA in Q4 and Q5, whereas 74\% participants in Q6 think absolute performance-price is very useful.
This indicates that absolute performance-price might be comparably attractive as relative performance-price for users.

However, we admit that in some application scenarios, the guarantee of absolute performance lower bound and price upper bound could be essential, and this does not conflict with relative SLAs (e.g., we can guarantee both relative performance-price and absolute performance and price bounds in the same SLA).
However, Absolute performance and price guarantees are challenging.
A viable solution may require the collection of massive performance metrics using non-invasive monitoring technologies such as eBPF~\cite{ebpf} and establishing an accurate correlation model between queries, resources, and input data using machine learning.


\section{PixelsDB}\label{sec:pixelsdb-design}

In this section, we present the design of PixelsDB and demonstrate how (relative) flexible SLA can be implemented through architectural design, and how SOS can be implemented on public cloud computing resources such as AWS Lambda and ECS.


\begin{figure*}[htbp]
    \centering
    \begin{minipage}{0.252\textwidth}
        \centering
        \includegraphics[width=\linewidth]{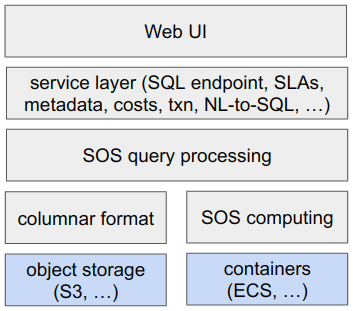}
        \vspace{-1em}
        \caption{Components overview.}
        \label{fig:overview}
    \end{minipage}\hspace{2.5em}\vspace{-0.5em}
    \begin{minipage}{0.63\textwidth}
        \centering
        \includegraphics[width=\linewidth]{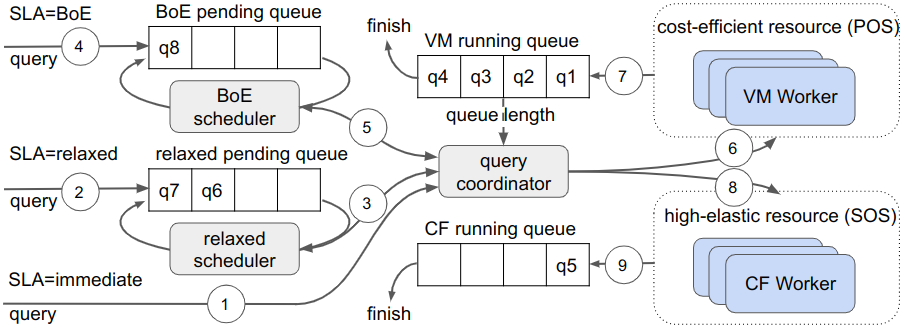}
        \vspace{-1em}
        \caption{Query processing solution to implement flexible SLA in PixelsDB.}
        \label{fig:query-scheduling-for-sla}
    \end{minipage}\vspace{-0em}
\end{figure*}

\subsection{Overview}\label{subsec:overview}

As shown in Figure~\ref{fig:overview}, PixelsDB is composed of four layers colored in grey.
Web UI is open-sourced at 
\blue{\href{https://github.com/pixelsdb/pixels-rover}{https://github.com/pixelsdb/pixels-rover}}.
Other layers are open-sourced in PixelsDB main repository.

\textbf{Web UI} provides a web user interface for data analytics.
It is demonstrated in~\cite{pixelsdb-demo}.
Despite basic functions such as schema browsing and query editing and execution, it also supports text-to-SQL, flexible SLA, and cost visibility.
Users can select a preferred service level when submitting queries.
They can also analyze the costs and performance of historical queries using brushing-and-linking.

\textbf{Service layer} provides the REST and gRPC APIs for clients (e.g., Web UI) to access the functionalities provided in PixelsDB.

\textbf{SOS query processing layer} is responsible for executing queries.
It coordinates the query tasks in the SOS computing environment.
PixelsDB currently supports hybrid query processing on computing resources with different cost-efficiency and elasticity (Figure~\ref{fig:query-scheduling-for-sla}).

\textbf{Columnar format} provides the columnar storage format that is used to store user data and intermediate materialized data.
The Pixels columnar format is proposed in~\cite{pixels-s3}.
It is optimized for a variety of storage infrastructures such as S3 and HDFS.

\textbf{SOS computing} provides the fundamental computing environment.
Currently, we use AWS Lambda~\cite{aws-lambda} as the SOS computing environment.
We have also integrated vHive~\cite{vhive-paper} into PixelsDB as an experimental SOS computing environment.
vHive is an open-source framework for serverless experimentation.
It is primarily composed of Kubernetes (K8s)~\cite{kubernetes}, Knative~\cite{knative}, Containerd~\cite{containerd}, and Firecracker~\cite{firecracker-paper}.


\textbf{Lessons learned:}
We find that Knative+K8s is not suitable for SOS computing due to their limitations in cold-start-latency stability, scaling mechanisms, and instance status management.
Hence, we are in the process of implementing SOS computing on public container services such as AWS ECS~\cite{aws-ecs} over EC2 and Fargate~\cite{aws-fargate}.

\subsection{Flexible Service Levels}\label{subsec:service-levels}

As shown in the left half of Figure~\ref{fig:query-scheduling-for-sla}, PixelsDB currently supports three service levels regarding query pending time.
We plan to implement SLAs regarding query execution time in the next step.

\textbf{(1) Immediate.}
At this service level, the service layer immediately submits the received query to the query coordinator (step \textcircled{1}), which decides whether to execute the query in the cost-efficient cluster 
or the high-elastic cluster.

\textbf{(2) Relaxed.} 
At this service level, the service layer enters the query into the \textit{relaxed pending queue} (step \textcircled{2}).
The query may be pending for at most 5 minutes (configurable).
The \textit{relaxed scheduler} keeps polling the queue.
If any query in the queue (1) can be executed in the cost-efficient cluster or (2) is approaching the pending time limit, the scheduler will dequeue and submit it to the query coordinator for execution (step \textcircled{3}).

\textbf{(3) Best-of-effort (BoE).}
At this service level, the service layer enters the query into the \textit{BoE pending queue} (step \textcircled{4}).
The \textit{BoE scheduler} keeps polling the queue.
Whenever the cost-efficient cluster is idle, the scheduler will dequeue a query and submit it to the query coordinator for execution (step \textcircled{5}).
There is no clear guarantee on the pending time.

The query coordinator decides whether to execute a query using the cost-efficient cluster or the high-elastic cluster.
This is discussed in Section~\ref{subsec:query-execution}.
Note that each service level only defines the upper bound of pending time.
Even a relaxed or BoE query may be executed immediately if the query coordinator permits.

\subsection{Query Processing}\label{subsec:query-execution}
The query processing layer in PixelsDB is based on Pixels-Turbo~\cite{pixels-turbo}.
It can execute queries using either cost-efficient resources (e.g., spot virtual machine like EC2 spot~\cite{AWS-EC2-spot}) or high-elastic resources (e.g., cloud function like AWS Lambda~\cite{aws-lambda}).
The purpose of hybrid query execution is to improve the overall cost-efficiency of a workload~\cite{pixels-turbo}.
Cost-efficient resources like spot VM are generally less elastic and require several minutes to scale in/out.
Whereas high-elastic resources like CF are more elastic (e.g., launching hundreds of instances in 1-2 seconds) but require 9-24x higher unit prices~\cite{pixels-turbo}.
Therefore, PixelsDB dynamically creates CF workers to process the workload spikes when the VMs are overloaded and can not scale out in time.
VM and CF are just two examples of heterogeneous cloud resources, and they are used in the current version of PixelsDB.
In fact, there are more types of resources with varying cost-efficiency and elasticity in public clouds~\cite{aws-eks,aws-fargate,aws-ecs}.

Currently, the query execution in the high-elastic resource (i.e., AWS Lambda) follows the SOS paradigm.
The query plan is compiled into multiple stages, with each stage executed in parallel by a set of Lambda instances.
Each Lambda instance has isolated computing resources (e.g., 6 vCPUs, 10GB memory, and around 800Mbps network bandwidth) and is only responsible for executing a task of a stage.
The query execution in the cost-efficient resource (i.e., AWS EC2) is currently implemented using Trino, which is an MPP query engine that follows the POS paradigm.
In the next stage of PixelsDB, we will remove Trino and replace Lambda by a dedicated SOS computing environment backed by cloud container services such as AWS EC2.
Thus, PixelsDB will become a pure SOS query engine.
We can eliminate the performance overhead introduced by SOS through the collaborative optimization of the query optimizer, the SOS query execution layer, and the SOS computing layer.
Then, the SOS computing layer will provide computing resource scheduling on heterogeneous computing resources.

To support the service levels in Section~\ref{subsec:service-levels}, as shown in the right half (after step \textcircled{5}) of Figure~\ref{fig:query-scheduling-for-sla}, the \textit{query coordinator} monitors the length of the \textit{VM running queue}~\footnote{The VM and CF running queues are for the queries running in the VM cluster and CF cluster, respectively. Queries enqueue when start running and dequeue when finished.} (i.e., query concurrency in the VM cluster).
Whenever a query arrives, the query coordinator decides whether to execute it in the VM (cost-efficient cluster) or the CF (high-elastic) cluster using one of the following policies.

\textbf{(1) Force.}
In this policy, the query coordinator forces the relaxed or BoE query to execute in the VM cluster (steps \textcircled{6} and \textcircled{7}), and it executes the immediate query in the CF cluster when the VM cluster is overloaded (steps \textcircled{8} and \textcircled{9}).
In this way, service level may directly decide the type of resource being used for a query.

\textbf{(2) Auto.}
In this policy, the query coordinator automatically decides to execute the query in the CF cluster when the VM cluster is overloaded, regardless of the query's service level.
In this way, the relaxed and BoE service levels indirectly affect resource consumption by mitigating workload spikes.

\textbf{Lessons learned:}
Currently, POS (based on Trino~\cite{trino-main-page}) is applied in the VM cluster.
We find (see Section~\ref{subsec:effects-of-scaling-paradigm}) that POS shares the cluster among many concurrent query tasks.
Thus, it increases task interference and resource competition, making it difficult to guarantee query performance and costs.
This is in line with the vision in Section~\ref{subsec:vision}.
Hence, we are developing a cost-efficient serverless computing framework for SOS based on containerized resources such as AWS ECS~\cite{aws-ecs} to replace the POS VM cluster.

\section{Evaluations}
In this section, we evaluate PixelsDB on near-real-world workloads generated by Cloud Analytics Benchmark (CAB)~\cite{cloud-analytics-benchmark}.
Besides confirming the SLAs are guaranteed, we analyze the effects of SLAs on performance and cost.

\subsection{Experimental Setup}
All the experiments are performed on AWS Lambda (CF) and an EC2 m5.8xlarge instance (VM) with 32 vCPUs and 128 GB memory.
The data is stored in Pixels columnar format~\cite{pixels-s3} in AWS S3.

\begin{figure}[tbp]
	\vspace{-0.5em}
	\begin{center}
		{\includegraphics[scale=0.3]{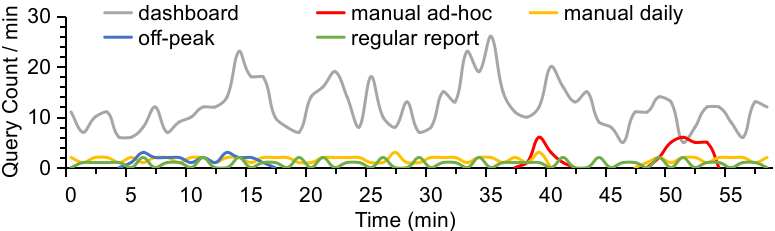}}
	\end{center}
	\vspace{-1em}
	\caption{Query stream of the workloads.}
    \label{fig:query-stream}
	\vspace{-0em}
\end{figure}

\textbf{Query stream.}
We use CAB~\cite{cloud-analytics-benchmark} to generate the workloads for evaluations.
CAB simulates the typical warehouse capacities and workload patterns in real-world cloud analytic workloads in Snowset~\cite{snowset}.
While CAB reuses the TPC-H data generator, it generates its own query workloads with five patterns regarding the query complexity and the timing of query arrivals.
As shown in Table~\ref{tab:benchmark}, we generate five datasets (databases) for the five workloads (one workload per pattern) using CAB.
In Table~\ref{tab:benchmark}, the fourth column is the number of queries in each workload, and the last column is the service level we assign to each workload.
The ratios in the last column represent the proportions of queries of different SLAs.
For the \textit{dashboard} workload, we use the relaxed SLA for 3/4 queries because we observe that most dashboards in real applications may be updated every few minutes, while the rest may be updated more frequently and should use the immediate SLA.
As shown in Figure~\ref{fig:query-stream}, we merge the five workloads into the timeline to create a unified query stream, thus simulating a real-world cloud data warehouse serving a variety of applications in an organization.
We submit the query stream to PixelsDB using concurrent clients.
For simplicity of discussion, in the rest of this section, we name the queries submitted with different SLAs (shown in Table~\ref{tab:benchmark}) as immediate, relaxed, and BoE queries.

\textbf{Evaluations.}
The design in Sections~\ref{subsec:service-levels} and~\ref{subsec:query-execution} guarantees the maximum query pending under each service level.
In the evaluations, we further verify the effects of flexible SLA on query costs and query execution time.
For that, we execute the query stream using the \textit{auto} or \textit{force} execution strategy with or without flexible SLA enabled (namely \textit{auto w/ SLA}, \textit{auto w/o SLA}, \textit{force w/ SLA}, and \textit{force w/o SLA}, respectively).
Then, we compare the cumulative query execution time and cost of the queries in each condition.
To disable flexible SLA, we force the service layer in PixelsDB to rewrite the SLA to immediate (the default behavior of traditional query service without flexible SLA) for all received queries.

\begin{figure}[tbp]
	\vspace{-0.5em}
	\begin{center}
		{\includegraphics[scale=0.275]{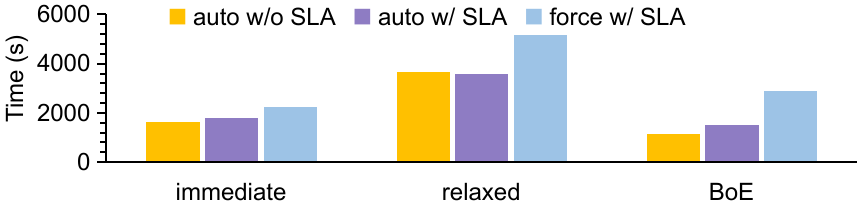}}
	\end{center}
	\vspace{-1em}
	\caption{Cumulative execution time of the queries submitted with different SLAs.}
    \label{fig:exp-perf}
	\vspace{-0em}
\end{figure}

\begin{figure}[tbp]
	\vspace{-0.5em}
	\begin{center}
		{\includegraphics[scale=0.275]{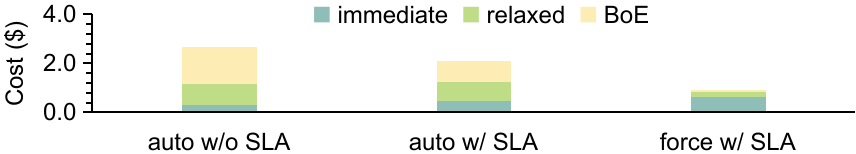}}
	\end{center}
	\vspace{-1em}
	\caption{Cumulative cost of the queries submitted with different SLAs.}
    \label{fig:exp-cost}
	\vspace{-0em}	
\end{figure}

The evaluation results are shown in Figure~\ref{fig:exp-perf} and Figure~\ref{fig:exp-cost}.
As \textit{auto w/o SLA} is actually equivalent to \textit{force w/o SLA} (i.e., all queries are immediately executed in VMs or CFs according to the load in VMs), we do not show \textit{force w/o SLA} in the evaluation results.

\begin{table}
    \centering
    \begin{tabular}{@{}ccccc@{}}
        \toprule
        \textbf{DB} & \textbf{Size (GB)} & \textbf{Workload Pattern} & \textbf{\#} & \textbf{SLAs} \\ \midrule
        $db_{1}$ & 10 & dashboard & 720 &  Rel/Imm=3/1  \\
        $db_{2}$ & 30 & manual ad-hoc & 34 & Imm   \\
        $db_{3}$ & 30 & manual daily & 87  & Imm/Rel=2/1   \\
        $db_{4}$ & 100 & off-peak & 22 & BoE   \\
        $db_{5}$ & 100 & regular report & 48 & Rel \\ \bottomrule
    \end{tabular}
    \caption{Datasets (databases) and workloads.}
    \label{tab:benchmark}
    \vspace{-0cm}
\end{table}

\subsection{Effects of SLAs on Execution Time}\label{subsec:effects-of-flexible-sla}
As shown in Figure~\ref{fig:exp-perf}, \textit{auto w/ SLA} and \textit{auto w/o SLA} have comparable cumulative query execution times for all categories of queries.
The BoE query execution time of \textit{auto w/ SLA} is slightly higher than that of \textit{auto w/o SLA}.
This is because, with \textit{auto w/ SLA}, more BoE queries are executed in the VM (POS).
BoE queries are big queries (on $db_4$).
SOS (CF) can dynamically allocate more resources to process these big queries.
Hence, BoE queries are faster in CFs.
For \textit{force w/ SLA}, the relaxed and BoE queries have higher cumulative execution times because these queries are forced to squeeze into the VM.
The immediate query execution time of \textit{force w/ SLA} is slightly higher because more immediate queries are executed in CFs.
Immediate queries are small queries (on $db_1$), and CF dynamically allocates fewer resources for these queries, making them slower.




\subsection{Effects of SLAs on Cost}\label{subsec:effects-of-scaling-paradigm}
As shown in Figure~\ref{fig:exp-cost}, \textit{auto w/ SLA} and \textit{force w/ SLA} have 22.2\% and 65.5\% lower total cumulative cost than \textit{auto w/o SLA}, respectively.
The cost reduction is contributed by the BoE and relaxed queries.
With \textit{auto w/ SLA} (or \textit{force w/ SLA}), these queries are more likely (or forced) to execute in the VM, which is more cost-efficient.
For immediate queries, the cost increases by 45.5\% and 99.9\% with \textit{auto w/ SLA} and \textit{force w/ SLA}, respectively.
This is because when flexible SLA is enabled, the VM is dominated by the relaxed and BoE queries; hence, the immediate queries are more likely to be executed in CF, which is more expensive.
However, compared with pure-CF query processing (the case in normal serverless query engines such as Starling~\cite{starling-paper} and Lambada~\cite{lambada-paper}), this cost is still 2x-3x lower.

\textbf{Takeaway:}
Using the auto execution policy, flexible SLA can reduce query costs without significantly affecting query performance.
Allowing SLA to directly determine the resources for query execution (i.e., \textit{force w/ SLA}) can further reduce query costs.
However, the current implementation of PixelsDB uses POS in the cost-efficient cluster (and uses SOS in the high-elastic cluster), leading to difficulty in controlling performance and cost.
In real applications, customers may also need SLAs regarding query execution time.
This requires improving the isolation and elasticity of resource management by applying a pure SOS query execution paradigm, as discussed in Section~\ref{subsec:vision}.
\section{Related Work}\label{sec:related}
\textbf{Cost Intelligence.}
Zhang et al. propose the blueprint for a cost-intelligent cloud data warehouse in~\cite{cost-intelligent-data-analysis}.
This work inspires our idea of performance SLAs.
However, \cite{cost-intelligent-data-analysis} targets a more general bi-optimization problem of performance and cost.
It mainly considers how to estimate query cost-efficiency and calculate the optimal resource deployment and settings from the perspective of query optimization.
Whereas this paper proposes a more concrete problem (flexible SLA) for serverless query processing based on application insights and discusses how to solve it from the perspectives of query execution and computing framework.
This would be complementary to the query optimization solution in~\cite{cost-intelligent-data-analysis}.

\textbf{Flexible Performance-price.}
Redshift-Serverless recently proposed an intelligent cluster scaling mechanism that allows customers to choose between higher performance and lower price by dragging a slider bar~\cite{redshift-intelligent-scaling}.
However, such a flexible performance-price choice is applied to the whole warehouse (i.e., virtual cluster), and it is not effective immediately.
When the customer adjusts the performance-price slider bar, the backend query engine tunes the scaling controller for the whole warehouse.
The scaling controller starts collecting query logs and may adjust resource allocation for the warehouse on a per-day basis.
Therefore, the flexible performance-price provided by Redshift-Serverless is not the flexible SLAs on a per-query basis discussed in this paper.
Flexible SLA requires dynamically allocating isolated resources to each query.
Compared with the POS execution paradigm in traditional MPP query engines, the SOS query execution paradigm has advantages in per-query resource allocation and isolation.

\textbf{SOS query processing.}
As discussed in Section~\ref{subsec:resource-scaling-paradigms}, Google has applied SOS query processing in BigQuery (Dremel~\cite{dremel-2020} and Borg~\cite{google-borg-paper}).
It confirms that SOS is suitable for efficient query processing.
We believe the inefficiency of CF-based SOS is caused by the limitations of CF (cloud function), but not SOS.
The main contribution of this paper is not to propose SOS query processing but to highlight its significance for solving flexible SLA and envision its popularity in serverless query processing.
As discussed in Sections~\ref{subsec:overview} and~\ref{subsec:query-execution}, we are developing the while-box serverless computing framework for SOS based on containerized computing services such as ECS over Fargate and EC2.

\section{Conclusion}\label{sec:conclusion}
In this paper, we first propose the flexible SLA problem for serverless processing based on the insights from real applications and the user study among database practitioners.
Then, we summarize the serverless query processing architectures into two categories: plan-oriented scaling (POS) and stage-oriented scaling (SOS).
Based on that, we envision the challenges and opportunities for flexible SLA and other innovative database research.
Finally, we present the design of PixelsDB.
In this system, we confirm the feasibility of implementing flexible SLA through dedicated architectural design and the significance of SOS for flexible SLA.


\bibliographystyle{IEEEtran}
\bibliography{my}

\end{document}